# Saturable absorption of free-electron laser radiation by graphite near the carbon K-edge

Lars Hoffmann[1,2,3], Sasawat Jamnuch[4], Craig P. Schwartz[2,5], Tobias Helk[6,7], Sumana L. Raj[1], Hikaru Mizuno[1,2], Riccardo Mincigrucci[8], Laura Foglia[8], Emiliano Principi[8], Richard J. Saykally[1,2], Walter S. Drisdell[2,9], Shervin Fatehi[10,‡], Tod A. Pascal[4,11,12,†], Michael Zuerch[1,2,3,13,*]

1. Department of Chemistry, University of California, Berkeley, California 94720, USA
2. Chemical Sciences Division, Lawrence Berkeley National Laboratory, Berkeley, California 94720, USA
3. Fritz Haber Institute of the Max Planck Society, 14195 Berlin, Germany
4. ATLAS Materials Science Laboratory, Department of Nano Engineering and Chemical Engineering, University of California, San Diego, La Jolla, California, 92023, USA
5. Nevada Extreme Conditions Laboratory, University of Nevada, Las Vegas, NV 89154, USA
6. Institute of Optics and Quantum Electronics, Abbe Center of Photonics, Friedrich-Schiller University, 07743 Jena, Germany
7. Helmholtz Institute Jena, 07743 Jena, Germany
8. Elettra-Sincrotrone Trieste S.C.p.A., Strada Statale 14, 34149 Trieste, Italy
9. Joint Center for Artificial Photosynthesis, Lawrence Berkeley National Laboratory, Berkeley, California 94720, USA
10. Department of Chemistry, The University of Texas Rio Grande Valley, Edinburg, Texas 78539, USA
11. Materials Science and Engineering, University of California San Diego, La Jolla, California, 92023, USA
12. Sustainable Power and Energy Center, University of California San Diego, La Jolla, California, 92023, USA
13. Materials Sciences Division, Lawrence Berkeley National Laboratory, Berkeley, California 94720, USA

**The interaction of intense light with matter gives rise to competing nonlinear responses that can dynamically change material properties. Prominent examples are saturable absorption (SA) and two-photon absorption (TPA), which dynamically increase and decrease the transmission of a sample depending on pulse intensity, respectively. The availability of intense soft X-ray pulses from free-electron lasers (FEL) has led to observations of SA and TPA in separate experiments, leaving open questions about the possible interplay between and relative strength of the two phenomena. Here, we systematically study both phenomena in one experiment by exposing graphite films to soft X-ray FEL pulses of varying intensity, with the FEL energy tuned to match carbon 1s to $\pi^*$ or 1s to $\sigma^*$ transitions. It is observed for lower intensities that the nonlinear contribution to the absorption is dominated by SA attributed to ground-state depletion; for larger intensities (>$10^{14}$ W/cm$^2$), TPA becomes more dominant. The relative strengths of the two phenomena depend in turn on the specific transition driven by the X-ray pulse. Both observations are consistent with our real-time electronic structure calculations. Our results reveal the competing contributions of distinct nonlinear material responses to spectroscopic signals measured in the X-ray regime, demonstrating an approach of general utility for interpreting FEL spectroscopies.**

Saturable absorption (SA) is a nonlinear optical response characterized by a reduction in the relative absorption of the sample—or, conversely, an increase in transmission—with an increase in the intensity of incident light. Although this effect has been known since the 1940s[1], direct experimental observation has been limited due to the high intensity required of the incident light, which was unachievable before the development of modern laser technology. The phenomenon has now been widely demonstrated and applied in the visible and infrared regions[2]. Moreover, saturable absorbers play a key role in passive mode-locking of femtosecond laser oscillators[3,4]. Saturable absorption is exquisitely sensitive to the electronic state of the material being probed. As such, SA is often used as a probe of dynamics and recovery following a pump-probe event, typically using a probe in the visible or infrared. It can be used to extract detailed information on photoexcited scattering and diffusion of charge carriers, as well as structural changes such as recovery following melting.[5,6]

As compared with studying SA in transitions between electronic valence states, which may exhibit significant hybridization, studying SA in transitions from the well-localized and minimally hybridized core to the valence states provides a more direct route to understanding the underlying processes. Such studies became possible with the advent of free-electron lasers (FEL), which combine very high intensities with high photon energies to enable excitations resonant with and capable of depleting a chosen ground state in the XUV to hard X-ray regions. At the FLASH FEL, saturable absorption of the L-shell transition of aluminum was observed with photon energies of 92 eV (13 nm).[7] Subsequent observations have been reported for the tin N-edge at 24 eV (52 nm) and for the iron K-edge at 7.1 keV (0.17 nm).[8,9] In these experiments, SA was attributed to high photon flux depleting the ground state, which led to X-ray-induced transparency. In similar experimental settings, by contrast, unexpectedly high absorption has been observed at the aluminum K-edge at 1560 eV (0.79 nm) and carbon K-edge at 285 eV (4.35 nm), which was attributed to two-photon absorption (TPA).[10,11] It was proposed by Stöhr and Scherz that X-ray induced transparency could be induced by stimulated elastic forward scattering[12]. Experimental support for this hypothesis was shown for Co/Pd multilayers, where a strong flux dependence on the resulting transmission and diffraction contrast was observed. As the flux increased, transmission increased and diffraction contrast decreased; this was ascribed to the stimulated effects, including stimulated emission[13]. The loss of diffraction was also observed in a separate set of observations on Co/Pd multilayers[14]. These contrasting observations suggest that a complete physical picture of the processes at play when high-intensity X-ray pulses interact with solid-state systems is still missing. In addition, because core-hole lifetimes are typically on the order of femtoseconds[15], comparable to the pulse duration of X-ray FELs, these are transient phenomena exhibiting a strong dependence on the pulse duration.

Here we employ femtosecond free-electron laser light to systematically study the intensity dependence of saturable absorption in graphite around the carbon K-edge, in the soft X-ray region. We target several transition energies to excite the carbon 1s core electrons into different valence states. Moreover, time-dependent density functional theory (TD-DFT) simulations are used to show that the experimental measurements contain signatures of both saturable absorption and two-photon absorption, with TPA becoming more dominant at higher intensities. To our knowledge, this work is the first to observe both regimes in one experiment and to disentangle the respective contributions.

**Experimental Details**

The experiments were performed at the EIS-TIMEX beamline of the FERMI free-electron laser.[16–18] The FEL beam first passed through the Photon Analysis, Delivery, and Reduction System (PADReS), which includes beam diagnostics and provides the incident pulse energy for every FEL shot. The FEL pulse (photon energy = 285.7–309.2 eV, pulse duration τ ≈ 25 fs FWHM, pulse energy $E_p$ = 4 – 18 µJ, spot size 12 × 12 µm$^2$) was focused on the sample (**Fig.**

**1a**). The transmitted beam was propagated through a 50 nm Ni filter to prevent camera saturation, then dispersed by a grating (HZB 1603 2, 1000 gr/mm) onto a CCD camera (Andor iKon-M SO). A single spectrum was recorded for each pulse. The samples were unsupported films of graphite with thickness of 80 nm. The films were raster scanned to probe a pristine spot with each shot. For each laser shot, comparing the recorded incident photon flux with the recorded photon flux post-sample results in a measure of sample transmission as a function of incident flux. The observed transmission is subsequently compared to the linear Beer–Lambert law. Use of the Beer–Lambert law assumes negligible reflectivity at normal incidence, which holds for carbon at 300 eV (refractive index n = 1).[19] The FEL photon energy is varied to probe π* (285.7 eV) and σ* (309.2 eV) regions from the 1s core state.

For a mechanistic understanding of the experiment, velocity-gauge real-time time-dependent density functional theory (VG-RTTDDFT) was employed[20,21] with a numerical atomic orbital basis set in order to propagate the electronic structure of graphite under an intense laser field. Exchange-correlation (XC)[22] effects are treated within the adiabatic local density approximation (ALDA) using the Perdew–Zunger LDA[23]. The carbon pseudopotential was generated by pseudoizing C:{1s,2p,3d} with the explicit inclusion of a 1s core hole. The C 2s state was then obtained as a higher energy solution of the atomic Schrödinger equation. The graphite primitive unit cell was subsequently propagated for 25 fs under a $\sin^2$-enveloped pulse centered at t = 12.5 fs, with incident pulse intensities ranging from $10^{10}$–$10^{14}$ W/cm$^2$. The calculations were performed for photon energies corresponding to experimental values for the π* and σ* regions.

**Results and Discussion**

The excitation scheme is illustrated in **Fig. 1a**, with the linear X-ray absorption spectrum of graphite[24] shown in **Fig. 1b** as a reference for the FEL energies used in this study. At the selected energies, the FEL excites the core electrons into either the π* orbitals or the σ* orbitals. The corresponding energy-level diagram is shown in **Fig. 1c**, indicating the respective transitions. The transmitted X-ray flux was measured as a function of varying FEL intensity, and it was observed that the transmission increases with the intensity in a strongly nonlinear fashion. **Figs. 2a,b** show the transmission through the sample at a range of intensities and for two specific energies. In **Fig. 2a,** probing 1s to π* transitions, the transmission change is lower than expected relative to the linear absorption behavior modeled by the Beer–Lambert law, i.e. sub-linear. This effect, also known as "reverse" saturable absorption, has been reported for the aluminum K-edge and carbon K-edge and was attributed to two-photon absorption.[10,11] In **Fig. 2b,** probing 1s to σ* transitions, the transmission is greater than expected from a linear response, i.e. super-linear. This nonlinear increase in transmission mirrors the previously reported observation of saturable absorption for the aluminum L-edge and iron K-edge.[7,9] In those cases, SA is caused by the finite core-hole lifetime, which at soft X-ray energies depends mainly on the rate of Auger decay. For graphite, the lifetime of the 1s hole was calculated to be 7 fs (Ref. [25]), which is of the same order of magnitude as the FEL pulse duration (~25 fs).

To gain additional insights into the observed trends, we turn our attention to first-principles calculations. In these calculations, the time evolution of the absorbed energy per unit input $E_{absorbed}(t) = E(t) - E(0)$ is evaluated at two different photon energies, representing the 1s to π* and σ* transitions. For the results of additional calculations at the pre-edge, see Supplemental Material **Fig. S1**.[26] We note that the energy deposited into the system is conserved. For π* transitions (**Fig. 3a**), a clear trend indicative of saturable absorption is visible in the range of $10^{11}$–$10^{14}$ W/cm$^2$, with the absorbed energy decreasing as the intensity increases. Transitions at and above an intensity of $10^{14}$ W/cm$^2$ do not uniformly follow this trend, sometimes showing increasing absorption. For the σ* transitions (**Fig. 3b**), similar behavior indicative of saturable absorption is observed for intensities of $10^{11}$–$10^{13}$ W/cm². Around $10^{14}$ W/cm², a drastically different behavior is observed in the calculation, with the absorption increasing substantially. The Fourier transform of the time-dependent current at $10^{14}$ W/cm$^2$ (**Fig. 3c,d**) shows an emerging signal at $2\omega$, indicating two-photon absorption (TPA). In addition, the overall absorption increases significantly for approximately 10 fs after excitation onset, then levels off after another 5–10 fs. The plateau suggests a core-depletion effect leading to two-photon excitations to very high energy states. To test this hypothesis, we evaluated the final energy of the system at t = 25 fs, i.e., the total energy absorbed from the X-ray pulse. The transmitted intensity inferred from the energy absorbed by the system was fit to a model allowing for contributions from saturable, non-saturable, and two-photon absorption:

$$T(I) = \exp\left(-\left[\frac{\alpha_0}{1 + I/I_{sat}} + \alpha_{NS} + \beta I\right]d\right). \qquad (1)$$

Here $T(I)$ is the transmission, and $\alpha_0$, $\alpha_{NS}$, and $\beta$ are constants related to saturable, non-saturable, and two-photon absorption, respectively, and $I_{sat}$ is a characteristic saturation intensity. The thickness $d$ is set to 0.68 nm, matching the thickness of the graphite primitive cell used in the simulation. As shown in **Figs. 2c,d**, this model (orange line) produces good fits to the calculated intensity data (solid blue circles) for both the 1s to π* and σ* transitions ($R^2$ = 0.98 and 0.96, respectively; see Supplemental Material **Table S1** for the parameter values).[26] If two-photon absorption is omitted (green line in **Figs. 2c,d**), however, a monotonic increase in transmission is observed. Our model indicates that the increase in transmission at relatively low intensity is purely due to saturable absorption. At higher intensities, ratios of the model constants (**Table S1**) show that saturable absorption is approximately twice as strong at the 1s to σ* transition compared to the 1s to π* transition, and two-photon absorption is approximately three times stronger.

    With these insights in hand, we can return to consideration of the experiment, and in particular, the sub-linear (TPA-dominated) and super-linear (SA-dominated) behavior of the transmission with increasing intensity at the respective 1s to π* and 1s to σ* transitions. To explain this difference, we first note that excitation into the σ* state at normal incidence is known to have a larger oscillator strength than excitation into the π* state.[27] While the samples used here are polycrystalline rather than single crystals, we expect that a similar reduction in the likelihood of initial core excitation into the π* limits the ground-state depletion that favors SA rather than TPA, thereby reducing the transmission relative to the linear model. At the 1s to σ* transition, by contrast, depletion of the 1s ground state is more likely, and SA can dominate in the measured intensity range. Second, we emphasize that **Figs. 2a,b** cover different intensity ranges. The experimental data for 1s to π* transitions in **Fig. 2a** span 1.9–3.3·$10^{13}$ W/cm$^2$, while those for 1s to σ* transition in **Fig. 2b** cover a more limited and lower intensity range, 1.3–2.2·$10^{13}$ W/cm$^2$. Although TPA is calculated to be more likely for the 1s to σ* transition than for the 1s to π* transition at comparable intensities, the observed difference in behavior can be attributed to the higher FEL intensities achieved during our measurements for π*.

    To summarize, in this work we report the complex interplay of saturable absorption and two-photon absorption of intense X-rays in graphite, taking advantage of the high intensities available at EIS-TIMEX to measure transmission near the carbon K-edge. In contrast to the mechanism proposed to explain previous observations of saturable absorption in graphite,[11] our calculations indicate that this trend of increasing transmission with increasing intensity originates in depletion of the core by the intense FEL pulse which persists until a regime is reached where TPA becomes dominant. Data collected at intensities of up to ~$10^{13}$ W/cm$^2$ exhibited a decrease or an increase in transmission with intensity, relative to a linear model, for the respective 1s to π* and 1s to σ* transitions. We attribute this behavior to different transition dipoles shifting the regime where TPA becomes dominant over SA at different

threshold intensities. Our calculations indicate that TPA will dominate for both photon energies at intensities greater than ~$10^{14}$ W/cm$^2$. Our experimental methods enable a detailed understanding of both linear and nonlinear processes that occur due to the absorption of intense radiation at X-ray energies and can readily be extended to other materials. These findings are relevant for correctly interpreting X-ray absorption and scattering data collected at high intensities.

**Acknowledgements:** Soft X-ray SHG measurements were conducted at the EIS-TIMEX beamline at FERMI. The research leading to these results has received funding from the European Community's Seventh Framework Program (FP7/2007-2013) under grant agreement nº 312284. M. Z., T. H., and L. H. acknowledge support by the Max Planck Society (Max Planck Research Group). M. Z. acknowledges support by the Federal Ministry of Education and Research (BMBF) under "Make our Planet Great Again – German Research Initiative" (Grant No. 57427209 "QUESTforENERGY") implemented by DAAD. S. F. acknowledges support for research in the UTRGV Department of Chemistry from Robert A. Welch Foundation Departmental Grant #BX-0048. S. L. R. and H. M. was supported by the U.S. Army Research Laboratory (ARL) and the U.S. Army Research Office (ARO) under Contracts/Grants No. W911NF-13-1-0483 and No. W911NF-17-1-0163. S. L. R. received a National Science Foundation Graduate Research Fellowship under Grant No. DGE 1106400. Any opinions, findings, and conclusions or recommendations expressed in this material are those of the author(s) and do not necessarily reflect the views of the National Science Foundation. W. S. D. acknowledges support from the Joint Center for Artificial Photosynthesis, a DOE Energy Innovation Hub, supported through the Office of Science of the U.S. Department of Energy, under Award No. DE-SC0004993. This research used resources of the National Energy Research Scientific Computing Center, a DOE Office of Science User Facility supported by the Office of Science of the U.S. Department of Energy under Contract No. DE-AC02-05CH11231. This work also used the Extreme Science and Engineering Discovery Environment (XSEDE), which is supported by National Science Foundation grant number ACI-1548562. M. Z. acknowledges funding by the W. M. Keck Foundation, and funding from Laboratory Directed Research and Development Program at Berkeley Lab (107573). M. Z. and T.A.P. acknowledge funding from the UC Office of the President within the Multicampus Research Programs and Initiatives (M21PL3263). The authors are grateful for discussion with Eric Neuscamman, Scott Garner and Joachim Stöhr.

**Author contributions**: L.H., T.H., H.M., S.L.R., R.M., L.F., E.P., C.P.S., W.S.D., S.F., and M.Z. conducted the experiments at the FERMI free-electron laser. L.H., C.P.S., W.S.D, S.F. and M.Z. analyzed and interpreted the experimental data. W.S.D., C.P.S., and M.Z. conceived the experiment. M.Z. supervised the project. T.A.P. directed the computational portions of the project, while S.J. performed all the calculations. L.H., C.P.S., S.F., T.A.P. and M.Z. wrote the manuscript with input from all authors.

**Competing Interests:** The authors have no competing interests.

**Correspondence:** Correspondence and requests for materials should be addressed to M. Zuerch (* mwz@berkeley.edu), Shervin Fatehi (‡ shervin.fatehi@utrgv.edu), and Tod A. Pascal († tpascal@ucsd.edu).

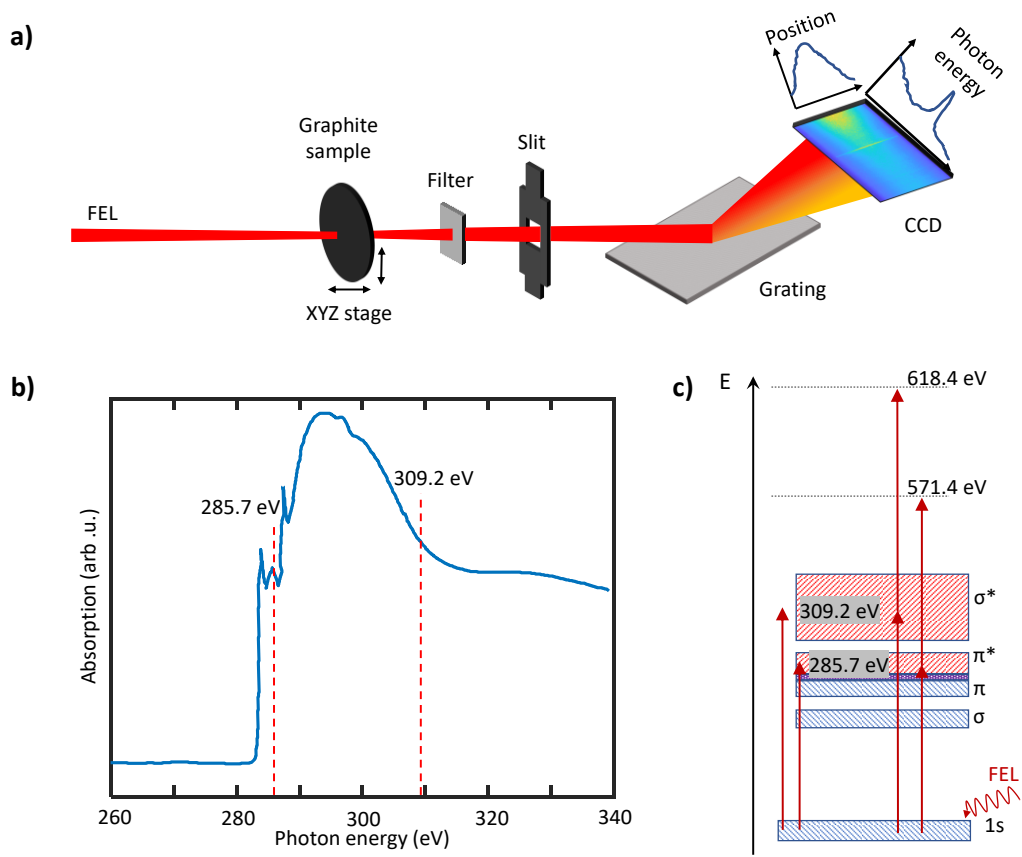

**Figure 1: Experimental setup and transitions in graphite**. a) Tunable FEL pulses were focused on the graphite sample. Ni metal foil was used to filter prevent camera saturation. An imaging spectrometer collected the transmitted X-ray light. To reference the incoming X-ray pulse intensity, a gas ionization–based intensity monitor was used downstream. b) X-ray absorption spectrum of a 500 nm graphite sample (data taken from Ref. 11). X-ray transmission measurements were conducted at discrete photon energies indicated with dashed lines. c) Energy level diagram of the absorption at the absorption edge in resonance with the π*-orbital (285.7 eV) and above the edge in resonance with the σ*-orbital (309.2 eV) as well as the respective two-photon absorptions.

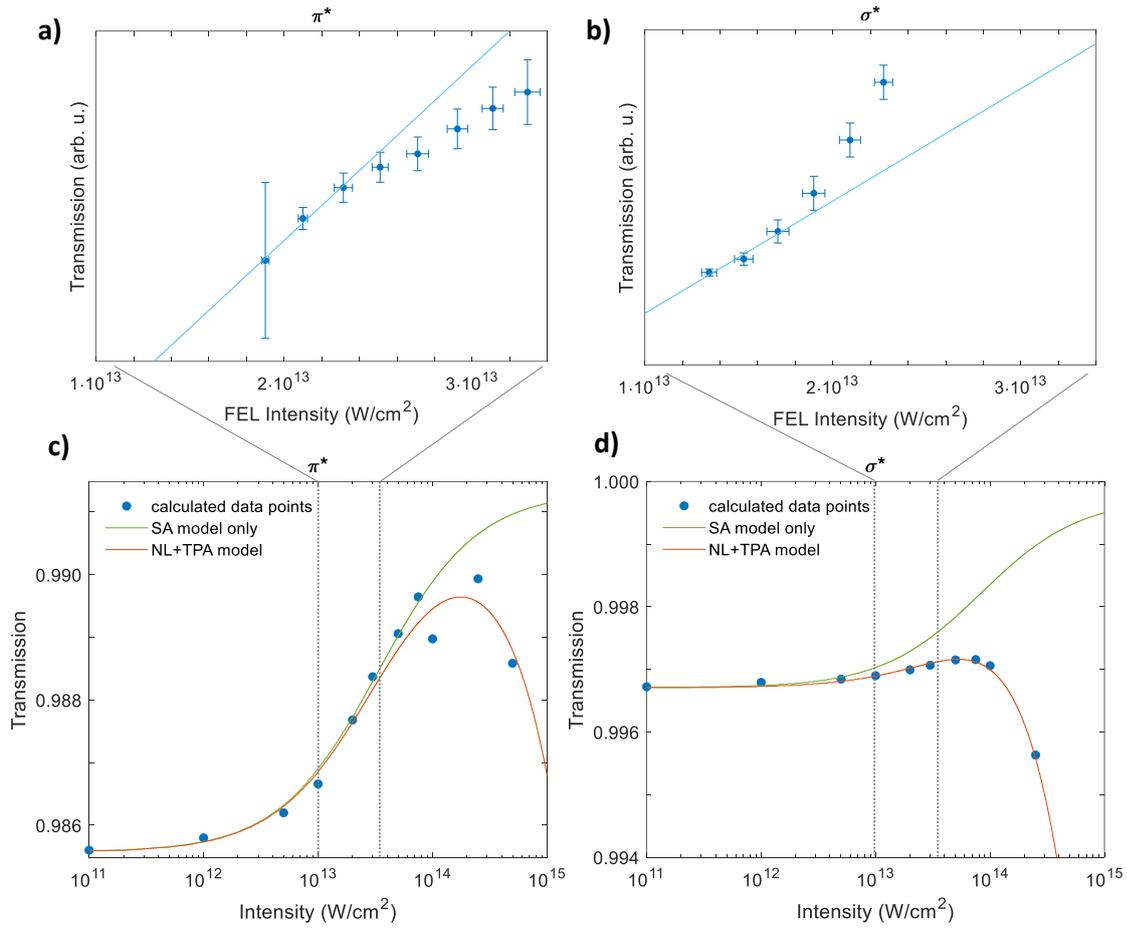

**Figure 2: Experimental and simulated transmission for resonant excitation from C 1s into π* and σ* states.** a) and b) Experimental normalized transmission for π* (285.7 eV) and σ* (309.2 eV) for FEL intensities in the region shown in c) and d) as dotted lines. Plot a) shows an increase in absorption while plot b) shows a decrease in absorption relative to the linear absorption model. Respective linear fits to the first three data points highlight the differing nonlinear behavior. c) and d) show the simulated transmission vs. intensity for π* and σ* fitted to the model in Eq. 1, including two-photon absorption (TPA, orange line). Comparison with a model neglecting TPA (green line) shows that TPA becomes dominant at high intensities. Gray dotted lines indicate the intensity range that was measured in the experiment.

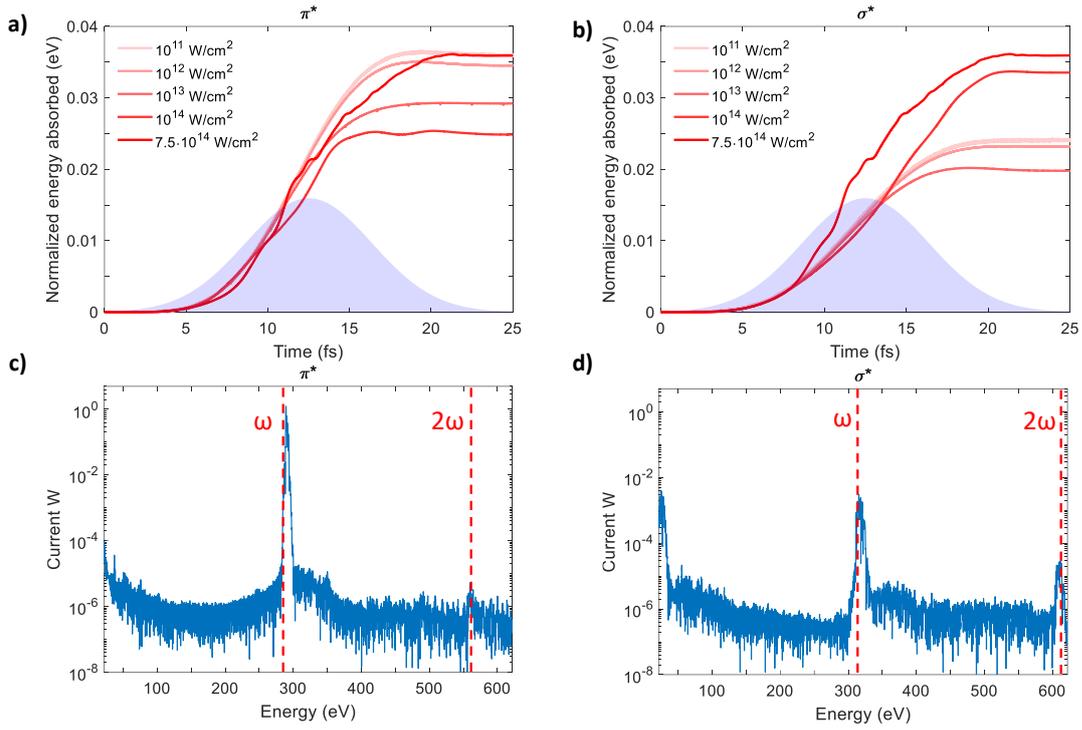

**Figure 3: Simulation of time-dependent absorption for resonant excitation from C 1s into π\* and σ\*.** a) and b) Time-dependent absorption for a transition C 1s → π\* (285.7 eV) and C 1s → σ\* (309.2 eV) for different X-ray intensities, respectively. The purple shaded area represents the envelope of the driving pulse used for the numerical simulation. Shown in c) and d) are the corresponding Fourier transforms of the time evolution of the current, indicating an additional component of the absorption at the highest intensities that stems from absorption of two photons of the FEL pulse 2ω.

# Saturable absorption of free-electron laser radiation by graphite near the carbon K-edge

Lars Hoffmann[1,2,3], Sasawat Jamnuch[4], Craig P. Schwartz[2,5], Tobias Helk[6,7], Sumana L. Raj[1], Hikaru Mizuno[1,2], Riccardo Mincigrucci[8], Laura Foglia[8], Emiliano Principi[8], Richard J. Saykally[1,2], Walter S. Drisdell[2,9], Shervin Fatehi[12,‡], Tod A. Pascal[4,11,12,†], Michael Zuerch[1,2,3,13,*]

1. Department of Chemistry, University of California, Berkeley, California 94720, USA
2. Chemical Sciences Division, Lawrence Berkeley National Laboratory, Berkeley, California 94720, USA
3. Fritz Haber Institute of the Max Planck Society, 14195 Berlin, Germany
4. ATLAS Materials Science Laboratory, Department of Nano Engineering and Chemical Engineering, University of California, San Diego, La Jolla, California, 92023, USA
5. Nevada Extreme Conditions Laboratory, University of Nevada, Las Vegas, NV 89154, USA
6. Institute of Optics and Quantum Electronics, Abbe Center of Photonics, Friedrich-Schiller University, 07743 Jena, Germany
7. Helmholtz Institute Jena, 07743 Jena, Germany
8. Elettra-Sincrotrone Trieste S.C.p.A., Strada Statale 14, 34149 Trieste, Italy
9. Joint Center for Artificial Photosynthesis, Lawrence Berkeley National Laboratory, Berkeley, California 94720, USA
10. Department of Chemistry, The University of Texas Rio Grande Valley, Edinburg, Texas 78539, USA
11. Materials Science and Engineering, University of California San Diego, La Jolla, California, 92023, USA
12. Sustainable Power and Energy Center, University of California San Diego, La Jolla, California, 92023, USA
13. Materials Sciences Division, Lawrence Berkeley National Laboratory, Berkeley, California 94720, USA

Correspondence and requests for materials should be addressed to M. Zuerch (*mwz@berkeley.edu), Shervin Fatehi (‡shervin.fatehi@utrgv.edu) and Tod A. Pascal (†tpascal@ucsd.edu).

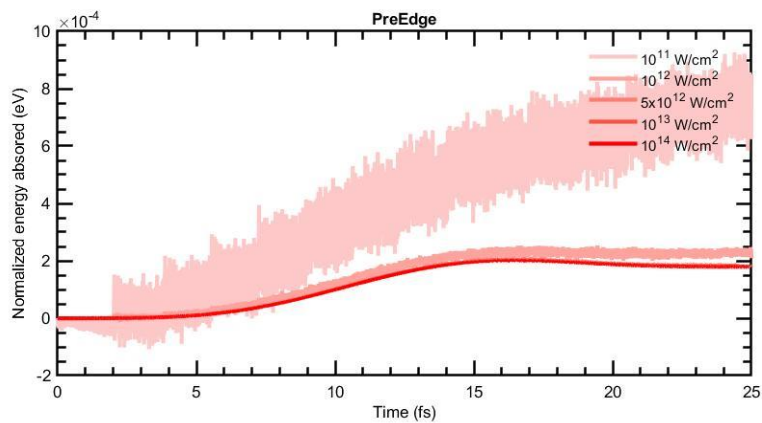

**Figure S1**: Calculated time evolution of energy absorbed at the pre-edge (261.6 eV). The absorption is low due to non-resonant conditions and appears largely independent of the pulse intensity. At lower intensities, the absorbed energy does not display a significant response, as the numerical noise within the simulation is on the same scale as the absorption.

**Table S1:** Model parameters used to fit transmission vs. intensity data from first-principles TD-DFT simulations.

|  | $\alpha_0$, nm$^{-1}$ | $I_{sat}$ (W/cm$^2$) | $\alpha_{NS}$, nm$^{-1}$ | $\beta$ (W/cm$^2$)$^{-1}$ nm$^{-1}$ |
|---|---|---|---|---|
| C1s -> $\pi^*$ | $5.8219 \cdot 10^{-3}$ | $5.70 \cdot 10^{15}$ | $8.7143 \cdot 10^{-3}$ | $4.3884 \cdot 10^{-7}$ |
| C1s -> $\sigma^*$ | $3.033 \cdot 10^{-3}$ | $1.56 \cdot 10^{16}$ | $2.6708 \cdot 10^{-4}$ | $1.3608 \cdot 10^{-6}$ |